\documentclass[twocolumn,showpacs,preprintnumbers,amsmath,amssymb]{revtex4}

\usepackage{graphicx}
\usepackage{dcolumn}
\usepackage{bm}
\usepackage{amsmath,dsfont}

\def\dAlembert{\vcenter {
    \hbox {\vrule height8pt width0.4pt depth0.0pt
           \vrule height8pt width7.2pt depth-7.6pt
           \vrule height8pt width0.4pt depth0.0pt
           \kern-8pt
           \vrule height0.4pt width8pt depth0.0pt
          \,}}}
\def\bA{{\bf A}}

\def\p{{\partial}}
\def\bx{{\bf x}}
\def\bY{{\bf Y}}
\def\by{{\bf y}}

\def\bR{{\mathds{R}}}

\def\bb{{\bf b}}
\def\ba{{\bf a}}

\def\bg{{\bf g}}

\def\bv{{\bf v}}
\def\tM{{\overline{M}}}
\def\bE{{\bf E}}
\def\bB{{\bf B}}
\def\bD{{\bf D}}
\def\bOmega{\mbox{\boldmath$\omega$}}
\def\bnabla{\mbox{\boldmath$\nabla$}}

\begin{document}

\preprint{0809.3128 [hep-th]}

\title{The geometry of Schr\"odinger symmetry
in  
 non-relativistic CFT}

\author{C. Duval}\email{duval-at-cpt.univ-mrs.fr}
\affiliation{Centre de Physique Th\'eorique, CNRS,
Luminy, Case 907\\
F-13288 Marseille Cedex 9 (France)} \altaffiliation{ UMR 6207 du
CNRS associ\'ee aux Universit\'es d'Aix-Marseille I et II et
Universit\'e du Sud Toulon-Var; Laboratoire affili\'e \`a la
FRUMAM-FR2291.}

\author{M.~Hassa\"\i ne}\email{hassaine-at-inst-mat.utalca.cl}
\affiliation{
Instituto de Matematica y F\'\i sica\\
Universidad de Talca\\
Casilla 747, Talca, Chile}

\author{P.~A.~Horv\'athy}\email{horvathy-at-lmpt.univ-tours.fr}
\affiliation{ Laboratoire de Math\'ematiques et de Physique
Th\'eorique, Universit\'e de Tours. Parc de Grandmont. F-37200 Tours
(France)}

\date{\today}

\begin{abstract}
The non-relativistic conformal
``Schr\"odinger'' symmetry of some gravity backgrounds proposed recently in the
AdS/CFT context, is explained in the ``Bargmann framework''.
The formalism incorporates the Equivalence Principle.
Newton-Hooke conformal symmetries, which are analogs of those
of   Schr\"odinger in the presence of a negative cosmological constant, are discussed in a similar way. Further examples include topologically massive gravity with negative cosmological constant and the Madelung hydrodynamical description.
\end{abstract}

\pacs{11.25.Tq, 03.75.Ss}

\maketitle

\section{Introduction}

Non-relativistic conformal transformations have initially been
discovered  as those space-time transformations that permute the
solutions of the free Schr\"odinger equation \cite{JackiwPT,Schr}. In $D+1$
dimensional non-relativistic space-time with position coordinates
$\by$ and time $t$ we have, in addition to the (one-parameter
centrally extended) Galilean generators,  also
\begin{equation}\begin{array}{llll}
(\by,t)&\to &(\by^*,t^*)=\left(\alpha\by,\alpha^2t\right)
&\hbox{\small dilatation}
\\[8pt]
(\by,t)&\to &(\by^*,t^*)=\left( \displaystyle\frac{\by}{1-\kappa t},
\displaystyle\frac{t}{1-\kappa t}\right)\qquad &\hbox{\small
expansions},
\end{array}
\label{NRconf}
\end{equation}
referred to as ``non-relativistic conformal transformations''. Added
to the Galilean symmetries provides us with the Schr\"odinger group.
Dilatations, expansions and time translations span an $o(2,1)\approx sl(2,\bR)$ subalgebra.

These rather mysterious extra symmetries have
 been identified as the isomorphisms of the structure of
 non-relativistic space-time \cite{DBKP, DGH}. For ``empty'' space, one gets, in particular, the (one-parameter centrally extended) Schr\"odinger group. Conformal transformations act as symmetries also when an
inverse-square potential  is introduced \cite{JackiwPT,
DeAlfaro}. In $D=3$ a
Dirac monopole can be included \cite{Jmonop,Horvathy:1983}, and in $D=2$ one can
have instead a magnetic vortex \cite{Jvortex}. Full Schr\"odinger
symmetry is restored for a matter field interacting with a
Chern-Simons gauge field \cite{JaPi,DHP}.
It can also be present in hydrodynamics \cite{hydro,
O'Raifeartaigh:2000mp}. See also \cite{Henkel:2003pu}.

Recently, the AdS/CFT correspondence has been extended to
non-relativistic field theory \cite{Bala,Son,other,Plyush}. The key point is to use the metric
\begin{equation}
\bar{g}_{\mu\nu}dx^{\mu}dx^{\nu}=
\frac{1}{r^2}\Big[
d\bx^2+dr^2+2dtds-\frac{dt^2}{r^2}
\Big]=\frac{1}{r^2}\,g_{\mu\nu}dx^{\mu}dx^{\nu}, \label{BSmetric}
\end{equation}
where  $\bx$ is an $d$-dimensional vector and $r$ an additional
coordinate.  The metric (\ref{BSmetric}) is  a
$d+3$ dimensional relativistic spacetime, conformally related to the
pp wave defined by $g_{\mu\nu}$ on the same manifold. The
interesting feature of the metric (\ref{BSmetric}) is that its {\it
isometries} are the  {\it conformal}
transformations of $d+1$ dimensional
non-relativistic space-time, with coordinates
$(\bx,t)$.

Below explain the construction and properties of this metric,
and illustrate it on some physical examples.

\section{Siklos spacetimes}

The metric (\ref{BSmetric}) belongs to the class of Siklos
spacetimes \cite{Siklos:1985,BaChGi,BeHa}, interpreted as exact gravitational waves traveling along $AdS$,
\begin{equation}
\bar{g}_{\mu\nu}dx^\mu dx^\nu
=\frac{1}{r^2}\left[d\bx^2+dr^2+2dtds
-F(\bx,r,t)dt^2\right].
\label{eq:AdSwave}
\end{equation}
 $\xi=\partial_s$ is a
 null Killing  vector. The metric
 (\ref{eq:AdSwave}) can  also be presented as
$\bar{g}_{\mu\nu}=\bar{g}_{\mu\nu}^{\mathrm{AdS}}
-{r^2F}\xi_{\mu}\xi_\nu,$
 generalizing the familiar Kerr-Schild transformation.  For $F=0$
 (\ref{eq:AdSwave}) reduces to the anti-de~Sitter metric.

The Einstein tensor of  (\ref{eq:AdSwave}) satisfies
\begin{eqnarray}\begin{array}{l}
\overline{G}_{\mu\nu}+\Lambda\bar{g}_{\mu\nu}=
\rho\,{\xi}_{\mu}\xi_{\nu},
\\[8pt]
\rho=\displaystyle\frac{r^4}{2}\Big(
\p^2_rF-\displaystyle\frac{d+1}{r}\p_rF+\bigtriangleup_{\bx}F\Big),
\end{array}
\label{eq:property}
\end{eqnarray}
where $\Lambda=-(d+1)(d+2)/2$. Hence, these spacetimes are solutions
of gravity with a negative cosmological constant, coupled to lightlike fluid. The only non-vanishing
component of the Einstein--de Sitter tensor is
$\bar{G}_{tt}+\Lambda\,\bar{g}_{tt}$.

The RHS of (\ref{eq:property}) is traceless, since
it is the energy-momentum tensor of some relativistic
fluid made of massless particles.  When
 $F$ satisfies the Siklos equation \cite{BaChGi},
\begin{equation}
\p^2_rF-\frac{d+1}{r}\p_rF+\bigtriangleup_{\bx}F=0,
\label{Sikloscond}
\end{equation}
then $\rho=0$, and (\ref{eq:AdSwave}) is the $AdS_{d+3}$ metric.

The effect of a conformal redefinition of the metric,
$g_{\mu\nu}\to\bar{g}_{\mu\nu}=\Omega^2g_{\mu\nu}$
has been studied by Brinkmann \cite{Brinkmann}.
Applied to our case we see that the Einstein equation of the pp wave
$g_{\mu\nu}$ in (1) goes over, for $\Omega=r^{-1}$,
 to that with negative cosmological constant,
appropriate for $\bar{g}_{\mu\nu}$.

\goodbreak
\section{Bargmann space and NR symmetries}

The best way to understand non-relativistic symmetries is  to use a
Kaluza-Klein-type framework \cite{DBKP,DGH}. The carrier space, $M$
(called ``Bargmann space''), is a $(D+1,1)$ dimensional spacetime,
endowed with a ``relativistic'' metric $g_{\mu\nu}$ and a
covariantly constant null vector, $\xi$. Such a metric can always be written as
\begin{equation}
g_{ij}dy^idy^j +2dt(ds+\bA\cdot d\by)-2Udt^2,
\label{gbmetric}
\end{equation}
where $g_{ij}=g_{ij}(\by,t)$ is a metric on $D$ dimensional space,
$\bA=\bA(\by,t)$ is a vector and $U=U(\by,t)$ is a scalar \cite{Brinkmann}. Then
$ \xi=\p_s$  is a covariantly constant vector.
In the sequel, we only consider the special case
$g_{ij}=\delta_{ij}$. We also assume, for simplicity, that $\bA$ and
$U$ do not depend explicitly on time.

Factoring out the curves generated by $\xi$ and represented by the
``vertical'' coordinate $s$ yields a $(D+1)$ dimensional manifold.
The relativistic metric does not induce a
metric on ordinary space-time
with coordinates $\by,t$. The contravariant
tensor $g^{\mu\nu}$ does project, however, together with the closed one-form $\xi_\mu$, endowing the quotient with a Newton-Cartan structure --
the structure of non-relativistic spacetime
\cite{DBKP}, with position $\by$ and non-relativistic time $t$.

Non-relativistic conformal symmetries are those conformal
transformations $f:~M\to M$
 which preserve the vertical vector $\xi$,
\begin{equation}
f^*g_{\mu\nu}=\Omega^2g_{\mu\nu}, \qquad f_*\xi=\xi \label{Bconf}
\end{equation}
where $f^*$  and $f_*$  denote  ``pull-back" and
``push-forward'', respectively.
The conformal factor, $\Omega$, is seen to depend only on $t$.

The origin of the terminology is that when $\bA=0$ and $U=0$ in
addition to $g_{ij}=\delta_{ij}$, Bargmann space is simply Minkowski
space written in light-cone coordinates. Then the $\xi$-preserving
isometries form precisely the one-parameter central extension of the
Galilei group called the Bargmann group;  those conformal
transformations which preserve $\xi$ span in turn the Schr\"odinger group. he latter acts on Bargmann space,
$(\by,t,s)\to(\by^*,t^*,s^*)$, according to
\begin{equation}
\left\{
\begin{array}{l}
{\bf y}^*=\displaystyle\frac{R{\bf y}+{\bf b}t+{\bf c}}{ft+g}
\\[8pt]
t^* =\displaystyle\frac{dt+ e}{ft+g}
\\[8pt]
s^* = {\displaystyle
s
+ \frac{f}{2}\frac{\left(R{\bf y} + {\bf b}t + {\bf c}\right)^2 }{ft + g}
-{\bf b}\cdot R{\bf y}
-\frac{t}{2}{\bf b}^2 + h,
}
\end{array}
\right.
\label{BargSchAc}
\end{equation}
where $R \in SO(D); {\bf b},{\bf c} \in {\bf R}^D; d,e,f,g,h \in {\bf R}$ and
$dg - ef = 1$. The corresponding conformal factor in (\ref{BargSchAc}) is
$
\Omega=(ft+g)^{-1}.
$
In particular,  $s^*=s$ for a dilatation, and
\begin{equation}
s^*=
s-\displaystyle\frac{\kappa|\by|^2}{2(1-\kappa t)}
\label{NRconfs}
\end{equation}
for an expansion.

The only time-independent potential which is consistent with the
non-relativistic conformal symmetries (\ref{NRconf})-(\ref{NRconfs})
is  homogeneous of order $(-2)$,
$U(\lambda\by)={\lambda}^{-2}U(\by)$. Such a potential clearly
reduces the Schr\"odinger symmetry to its $SL(2,\bR)$ subgroup
(plus, perhaps, other residual symmetries, depending on the concrete
form of $U(\by)$).

Returning to the general case (\ref{gbmetric}), we remind the Reader that
non-relativistic mechanics can be discussed by considering null
geodesics \cite{DBKP,DGH}.
Turning to quantum mechanics,
``ordinary'' wave functions lift to Bargmann space as functions, $\psi$, that are equivariant with respect to vertical translations, viz.,
\begin{equation}
\xi^\mu\partial_\mu\psi=im\psi,
\label{equivar}
\end{equation}
where $m$ is some real constant, namely the Galilean mass. For such a function
$\Psi=e^{-ims}\psi(\by,t,s)$ is $s$-independent, $\Psi=\Psi(\by,t)$,
and can therefore be identified with an ``ordinary'' wave function.
Including the ``vertical'' coordinate incorporates the gauge degree
of freedom --- just like
for ``monopoles without string''.

Let us now consider the massless wave equation
\begin{equation}
\left[\dAlembert-\frac{D}{4(D+1)}R\right]\psi=0, \label{Mlaplace}
\end{equation}
where $\dAlembert$ is the Laplace-Beltrami operator of the Bargmann
metric. The inclusion of the scalar curvature, $R$,
insures the invariance of
(\ref{Mlaplace}) under conformal
rescalings $g_{\mu\nu}\to{}\bar{g}_{\mu\nu}=
\Omega^2{}g_{\mu\nu}$, with
$\psi\to\psi^*=\Omega^{D/2}\psi$. The equations
(\ref{equivar}) and (\ref{Mlaplace}) yield the covariant
Schr\"odinger equation on Newton-Cartan space-time
\cite{DBKP}. In the special case $g_{ij}=\delta_{ij}$ and $\bA=0$, the ordinary Schr\"odinger equation with potential $U$ is recovered,
\begin{equation}
i\p_t\Psi=\left[-\frac{\bigtriangleup}{2m}+U
\right]\Psi,
\label{simpSchreq}
\end{equation}
where $\bigtriangleup\equiv \bigtriangleup_{\by}$ is the usual
 $D$-dimensional Laplacian.

The natural action of $\xi$-preserving conformal transformation of $M$ induces one acting on an  ordinary wave function,
\begin{eqnarray}
\Psi\to \Psi^*&=&\Omega^{D/2}e^{im(s^*-s)}\Psi(\by^*,t^*).
\label{trimplem}
\end{eqnarray}
Owing to masslessness,
eqn. (\ref{Mlaplace}), and hence also (\ref{genSchreq}), are
invariant w.r.t. such transformations.
In the free case we get, in particular, the Schr\"odinger symmetry.

\section{The Equivalence Principle}

Let us now return to the general case, (\ref{gbmetric}) with nonvanishing $\bA$ and $U$.
For an equivariant function, (\ref{equivar}), our wave equation
(\ref{Mlaplace}) reduces to
\begin{equation}
i\p_t\Psi=\left[-\frac{1}{2m}\bD^2+U
\right]\Psi, \quad \bD=\bnabla -im\bA.
\label{genSchreq}
\end{equation}
This provides us with a natural interpretation
of the components of the metric (\ref{gbmetric})~:
$\bA$ and $U$ represent the vector  and scalar potential of an `electromagnetic' field,
\begin{equation}
\bE=-\bnabla  U,
\qquad
\bB=\bnabla \times\bA.
\label{efield}
\end{equation}
(\ref{genSchreq}) is plainly the minimally coupled Schr\"odinger
equation in this ``electromagnetic'' field.
Note, however, that the coupling constant here is the mass, $m$,  not the electric charge.

Consistently with \cite{Jmonop,Jvortex},
the $SL(2,\bR)$ survives the inclusion of a Dirac
monopole \footnote{For a monopole,
$ds+\bA\cdot d\by$ is a $U(1)$ connection form
on the Hopf bundle.}, or of an Aharonov-Bohm vector potential
into the components of the metric (\ref{gbmetric}) \cite{DGH,DHP2}.

The components in (\ref{gbmetric}) admit, however, another interpretation~:
$\bA$ and $U$ can also be viewed as associated with
{\it non-inertial coordinates}.
Let us explain this using an example.
Consider the metric,
\begin{equation}
d\by\strut^2+2dtds+2({\bg}\cdot\by)dt^2,
\label{ugfield}
\end{equation}
of a uniform gravitational field, ${\bg}$.
Let us now
switch to an uniformly accelerating coordinate system,
which amounts to using accelerated coordinates,
\begin{equation}
\bY = \by-\frac{1}{2}\bg t^2,
\quad
T = t,
\quad
S=s+(\bg\cdot\by)t-\frac{1}{6}\bg^2 t^3.
\label{accelcoord}
\end{equation}
When expressed in these coordinates, the metric (\ref{ugfield}) is
simply  that of a free particle,
\begin{equation}
d\bY^2+2dTdS.
\label{freemetric}
\end{equation}
Thus, the inertial forces compensate the uniform
gravitational field~: a particle in Einstein's  freely falling lift
is free \cite{Einstein}.

Conversely, one can start with a free system, (\ref{freemetric}),
and switch to coordinates in a system  which rotates
with angular velocity
$\bOmega=\omega\,\hat{\bf Z}$ around the axis perpendicular to
the $X-Y$ plane,
\begin{equation}
\left(\begin{array}{l}X\\[4pt] Y\end{array}\right)
=
\left(\begin{array}{c}
\cos\omega t\,{x}+\sin\omega t\,{y}
\\[4pt]
-\sin\omega t\,{x}+\cos\omega t\,{y}
\end{array}\right).
\end{equation}
Completed with
$T={t}$ and $S={s}$ and dropping the coordinate $Z$,
the free metric (\ref{freemetric}) reads, in these new terms,
\begin{equation}
d\by\strut^2
+2d\strut{t}\big[ds+(\bOmega\times\by)\cdot d\by\big]
+\omega^2\by^{2}
\,d\strut{t}\strut^2
\end{equation}
($\by=\big({x},{y}\big)$).  Switching to a rotating
coordinate system amounts, therefore, to generating
a `magnetic' field twice the angular momentum,
$\bB=2\bOmega$,
and an `electric' field representing the centrifugal force,
\begin{equation}
\fbox{$
\begin{array}{lll}
\hbox{electric charge}\ e
&\longleftrightarrow\quad
&\hbox{mass}\ m
\\[4pt]
\hbox{vector potential}\ \bA\quad
&\longleftrightarrow
&\hbox{Coriolis potential}\
2{\bOmega}\times\by
\\[4pt]
\hbox{scalar potential}\ U
&\longleftrightarrow
&\hbox{centrifugal potential}\
\frac{1}{2}\bOmega^2\by^{2}
\end{array}
$}
\label{correspondence}
\end{equation}

We conclude that the ``Bargmann'' framework incorporates
Einstein's {\it Equivalence Principle} into non-relativistic
physics,
in that it allows
to use {\it any} coordinate system.

The correspondence (\ref{correspondence})
has been used in explaining the
Sagnac effect, [an Aharonov-Bohm type interference
experiment on a turntable \cite{Sagnac}].
It has yet another, rather striking, consequence. Let us
rotate a bulk of Type II superconductor. According to
(\ref{correspondence}), this should
be equivalent to putting it into a combined electric and
magnetic field.
But Type II superconductors react to a (sufficiently weak)
magnetic field by {\it expelling it} (Meissner effect) -- and this should also happen
when rotating it.
This  has actually been experimentally observed by Zimmermann and Mercereau \cite{ZM}~:
rotating a superconductor generates in fact
an electromagnetic field called the {\it London moment} \cite{LON},
which precisely compensates for the inertial forces \cite{HILD}
-- this is the {\it `` inertial Meissner effect''}.

One could also start with slowly rotating our
Type II superconductor. Initially, the London moment
excludes the flux. Passing a critical value of the angular velocity, however, it should
create ``Abrikosov'' vortices.

The experiment can also be performed using by rotating a superfluid.
Then one should again create vortices -- which have been indeed
observed \cite{Varoquaux}.

In what follows, we restrict our attention to $\bA=0$.

\section{Properties of the metric (\ref{BSmetric})}

Let us now turn to the metric (\ref{BSmetric}). The
bracketed part, $g_{\mu\nu}$, is a Bargmann space with $D=d+1$. Let
us call $\tM$ the same manifold, endowed with the rescaled
metric $\bar{g}_{\mu\nu}$. One of the coordinates, denoted by $r$,
has been distinguished, $\by=(\bx,r)$.
The potential term, $U=r^{-2}$, breaks the
full, $D$-dimensional Schr\"odinger symmetry to $o(2,1)$. It plainly
has also the $d=(D-1)$ dimensional Schr\"odinger symmetry of
$(\bx,t)$ space.

 Special conformal transformations [expansions] act
 in particular,  as
$(\bx,r,t,s)\to (\bx^*,r^*,t^*,s^*)$,
\begin{eqnarray}
(\bx^*,r^*,t^*,s^*)&=&\left( \Omega\bx,\Omega r,\Omega t,
s-\Omega\displaystyle\frac{\kappa|\by|^2}{2} \right),
\\[6pt]
\Omega&=& \displaystyle\frac{1}{1-\kappa t}.
\label{xrexp}
\end{eqnarray}
Similar formulae hold for an arbitrary conformal transformation
\footnote{The action
(\ref{xrexp}) is natural when $\bx$ and $r$ are viewed
as parts of $\by$. It also corresponds to viewing the $d=D-1$
dimensional Schr\"odinger group as a subgroup of the $D$-dimensional one, It is not the only implementation, though~:
we could have also extended it by having it act trivially,
i.e., leaving $r$ invariant. Then $\by^2=\bx^2+r^2$
would be replaced by $\bx^2$.
Such an action would again be
 an isometry of the rescaled metric.
It seems that this is not the action people are interested
in, however.}. For any of them,
the bracketed part in
(\ref{BSmetric}),  $g_{\mu\nu}$, gets multiplied by the conformal factor
$\Omega^2$; but this is exactly compensated for by the
action on the pre-factor $r^{-2}$,
leaving us with an isometry of  the full metric
$\bar{g}_{\mu\nu}$.
Hence, passing from $g_{\mu\nu}$ to $\bar{g}_{\mu\nu}$
converts conformal Bargmann transformations into isometries.

The massless wave equation (\ref{Mlaplace}) on $D=d+1$ dimensional
Bargmann space, reduced to (\ref{genSchreq}), reads now
\begin{equation}
\left[i\p_t+\frac{\bigtriangleup}{2m}-\frac{m}{2r^2} \right]\Psi=0,
\label{Msch}
\end{equation}
where $\bigtriangleup\equiv\bigtriangleup_{\by}$ is the Laplacian of
$D=d+1$ dimensional space
with coordinate $\by=(\bx,r)$ \footnote{If we argue that only $\bx$ is ``physical'' and the coordinate $r$ has been added as an auxiliary variable,
we should require
that the wave function be independent of $r$. Then
the Laplacian in (\ref{Msch}) reduces to
$\bigtriangleup_{\bx}$.}.
Its symmetry w.r.t. non-relativistic conformal transformations is
obtained when the latter are implemented according to
(\ref{trimplem}).

What happens when we trade the pp-wave $g_{\mu\nu}$ for
$\bar{g}_{\mu\nu}$~? By analogy with (\ref{Mlaplace}) we postulate,
\begin{equation}
\left[\overline{\dAlembert}-\frac{d+1}{4(d+2)}\overline{R}\right]\phi=0,
\label{tMlaplace}
\end{equation}
where $\phi$ is equivariant, (\ref{equivar}), and $\overline{R}$ is
the curvature of $\tM$. The curvature of the rescaled metric does
not vanish, but is rather $\overline{R}=-(d+3)(d+2)$. The reduced
equation reads, therefore,
\begin{eqnarray}
\left[ i\partial_t+\frac{\bigtriangleup}{2m}
-\frac{d+1}{2mr}\partial_{{r}} -\frac{M^2}{2mr^2}\right]{\Phi}=0,
\label{modschro}
\end{eqnarray}
where
\begin{eqnarray}
M^2=m^2-\frac{(d+1)(d+3)}{4(d+2)^2}. \label{M}
\end{eqnarray}

In the $\tM$ framework the conformal transformations are implemented
without the conformal factor, $\phi\to
\phi^*=\phi(\bx^*,r^*,t^*,s^*)$, i.e.,
\begin{equation}
\Phi^*(\bx,r,t)=e^{im(s^*-s)}\Phi(\bx^*,r^*,t^*). \label{timp}
\end{equation}
The conformal factor has been absorbed into the new terms in the
modified equation (\ref{modschro}). Alternatively, direct
calculation shows that equation (\ref{Msch}) becomes
(\ref{modschro}) under the redefinition,
$\Psi=r^{-\frac{d+1}{2}}\Phi.$ The weight
$-(d+1)/2$ in the exponent is precisely the one
necessary to assure the conformal invariance of
(\ref{tMlaplace}).

As seen before, a non-relativistic conformal transformation acts on
$(\tM,\bar{g}_{\mu\nu})$ by isometries. Implementing them according
to (\ref{timp}) yields, once again, symmetries. Does we have more
conformal transformations than just the mere isometries~? The answer
is negative: a tedious calculation whose details will not be reproduced here shows, in fact, that any
$\xi$-preserving conformal transformation of the $\bar{g}_{\mu\nu}$
metric is necessarily an isometry.

\section{Conformal extension of the Newton-Hooke group}

In the same spirit, we can consider the Newton-Hooke spacetime, which
can be obtained as the limit  of a relativistic spacetime with
cosmological constant $\Lambda$  when $c^2\Lambda/3\to\tau^{-2}=$
const. \cite{Gibbons:2003rv}.
The extended space,
\begin{equation}
ds^2=d\by^2+2dtds-\frac{|\by|^2}{\tau^2}dt^2.
\label{NHst}
\end{equation}
is a homogenous plane wave. It is
 is again a Bargmann space, and
 plays a role analogous
to that of Minkowski space for the Galilei group.

The  isometries of (\ref{NHst}) which preserve
$\xi=\p_s$ form the Newton-Hooke group. The latter act on $D+1$
non-relativistic space-time, $(\by,t)$, obtained by factoring out
the vertical direction $s$, according to
$(t,\by)\to (t^*,\by^*)$,
\begin{eqnarray}
\left\lbrace
\begin{array}{lll}
t^*&=&t+\epsilon,
\\[2pt]
\by^*&=&{\cal
R}\by+\ba\cos(\frac{t}{\tau})+\bb\,\tau\sin(\frac{t}{\tau}),
\end{array}
\right. \label{NHt}
\end{eqnarray}
where ${\cal R}$ is an $SO(D)$ matrix, $\epsilon\in\mathds{R}$,
$\ba,\bb\in\mathds{R}^{D}$ are parameters
associated with spatial rotations, time translation, pseudo
translations and pseudo Galileo boosts, respectively.

Assuming equivariance with unit mass $m=1$, the wave equation
(\ref{Mlaplace}) becomes  the Schr\"odinger equation of a harmonic oscillator,
\begin{eqnarray}
\left[i\partial_t+\frac{1}{2}\bigtriangleup-\frac{|\by|^2}{2\tau^2}\right]\Phi=0,
\label{sl}
\end{eqnarray}
which reduces to the free Schr\"odinger equation when
$\tau\to\infty$.

Newton-Hooke transformations act as symmetries. This can also be seen
directly, using that their natural implementation on Bargmann space
reads, when expressed on non-relativistic objects,
\begin{eqnarray}
\Phi^*(t,\by)=\Phi(t^*,\by^*)\,e^{i\Theta(t,\by)}, \label{impl}
\end{eqnarray}
where $\Theta(t,\by)$  is
\begin{eqnarray}
\frac{1}{\tau}(\ba\cdot\by) \sin(\frac{t}{\tau})+\frac{\ba^2}{4\tau}
\sin(\frac{2t}{\tau})-(\bb\cdot\by)
\cos(\frac{t}{\tau})-\frac{\bb^2\tau}{4} \sin(\frac{2t}{\tau}).
\nonumber
\end{eqnarray}

Beyond Newton-Hooke transformations, equation (\ref{sl}) admits
two further symmetries, analogous to dilatations and expansions. The first of these is
\begin{eqnarray}
\left\lbrace
\begin{array}{l}
t^*=\tau\arctan\left(\alpha^2\tan\left(\frac{t}{\tau}\right)\right),
\\[16pt]
\by^*=\displaystyle{\frac{\alpha\,\cos\left[\arctan\left(\alpha^2\tan\left(\frac{t}{\tau}\right)\right)\right]}
{\cos\left(\frac{t}{\tau}\right)}}\,\by,
\end{array}
\right. \label{NHct1}
\end{eqnarray}
where $\alpha$ is the parameter of the transformation.
We also have an analog of expansions,
\begin{eqnarray}
\left\lbrace
\begin{array}{l}
t^*=\tau\arctan\Big(\frac{\tan\left(\frac{t}{\tau}\right)}{1-\kappa\tau\tan\left(\frac{t}{\tau}\right)}\Big),
\\[16pt]
\by^*=\displaystyle{\frac{\cos\Big[\arctan\Big(\frac{\tan\left(\frac{t}{\tau}\right)}{1-\kappa\tau\tan\left(\frac{t}{\tau}\right)}\Big)
\Big]}{\cos\left(\frac{t}{\tau}\right)(1-\kappa\tau\tan\left(\frac{t}{\tau}\right))}}\,\by.
\end{array}
\right. \label{NHexpansion}
\end{eqnarray}

These transformations are implemented by involving
both phase and ``conformal" factors,
\begin{eqnarray}
\Phi^*(t,\by)=\Omega_{}^{D/2}\,\Phi(t^*,\by^*)\,e^{i\Theta(t,\by)},
\label{implct1}
\end{eqnarray}
where
\begin{eqnarray}
\Omega_{}^2(t)=\alpha^2\frac{1+\tan\left(\frac{t}{\tau}\right)^2}{1+\alpha^4\tan\left(\frac{t}{\tau}\right)^2},
 \label{NHOmega}
\end{eqnarray}
for a dilatation and
\begin{eqnarray}
\Omega_{}^2(t)=\frac{1+\tan\left(\frac{t}{\tau}\right)^2}{1+(1+\kappa^2\tau^2)\tan\left(\frac{t}{\tau}\right)^2
-2\kappa\tau\tan\left(\frac{t}{\tau}\right)}
\label{NHOmegaexpansion}
\end{eqnarray}
for an expansion.
The phase change for the dilatation reads
\begin{eqnarray}
\Theta_{}(t,\by)=\frac{(\alpha^4-1)\,\by^2\,\sin(2t/\tau)}
{2\tau\left[1+\cos(2t/\tau)+\alpha^4(1-\cos(2t/\tau))\right]},
\end{eqnarray}
while for the expansion the expression is rather complicated, and is not reproduced here.
 A tedious computation shows that  the transformations (\ref{NHct1})
and (\ref{NHexpansion}) act as symmetries for (\ref{sl}).
 Note for further reference that, in both cases,
 $\by^*=\Omega\,\by$.

 In the limit $\tau\to\infty$, these symmetries become precisely the
dilatation and the expansion of the Schr\"odinger group.

Not surprisingly, these ``conformal'' transformations lift to
genuinely conformal transformation of the Bargmann metric
(\ref{NHst}). When completed with
$s^*=s+\Theta$, they become
conformal transformation of the Bargmann space with conformal
factor $\Omega$.

Bargmann conformal transformations can, again, be converted into
isometries following the same strategy as before. We detach one
coordinate, i.e. we write $\by=(\bx,r)$ and rescale the metric
(\ref{NHst}) as
\begin{equation}
\label{eq:ga}
d\bar{s}^2=\frac{1}{r^2}\left[
d\bx^2+dr^2+2dtds-\frac{(\bx^2+r^2)}{\tau^2}dt^2\right].
\end{equation}
Then the change of the bracketed quantity under a conformal
transformation is, once again, compensated by the change of $r^2$.

The metric (\ref{eq:ga}) is that of $AdS$ as it should, since the
function in front of $dt^2$ trivially satisfies (\ref{Sikloscond}).

The simplest way to understand the origin of Newton-Hooke
symmetries is  to realize that they are, in fact, ``imported'' Galilean symmetries
 \cite{JaPi,NiedererOsc,Burdet:1984xt,Jackiw:1991ns,
 DHP2}. Putting $\omega=\tau^{-2}$,
the time-dependent dilatation,
\begin{eqnarray}
\by\to\bY=\frac{1}{\cos\omega t}\,\by{},
\quad
t\to T=\frac{\tan\omega t}{\omega}\ ,
\label{osctr}
\end{eqnarray}
implemented as
\begin{eqnarray}
\Psi(\by,t)&=&\frac{1}{(\cos\omega t)^{D/2}}e^{-i\frac{\omega}{2}\by^2\tan\omega t} \Phi(\bY,T),
\label{osctrimp}
\end{eqnarray}
transforms the Newton-Hooke (in fact, oscillator) background problem into a ``free'' one.
When completed with
\begin{eqnarray}
S=s-\displaystyle\frac{\omega \by^2}{2}\tan\omega t,
\label{oscfree}
\end{eqnarray}
the Newton-Hooke Bargmann metric (\ref{NHst})  is carried by (\ref{osctr})-(\ref{oscfree})
Bargmann-conformally into the free form,
(\ref{freemetric})
\footnote{The transformation (\ref{osctrimp})
can be extended to uniform magnetic and electric field,
and a curl-free ``Aharonov-Bohm'' potential \cite{DHP2}.  The map (\ref{accelcoord}) does the same for the linear force field.}.

The conformal factor is $\Omega(t)=|\cos\omega t|^{-1}$.
Accordingly,
this transformation  maps  (\ref{eq:ga}) isometrically to AdS space.

\section{Topological gravity}

At last, the metric (\ref{eq:AdSwave})
is also encountered in
topologically massive gravity  in $(2+1)$ dimensions \cite{Deser:1981wh} with
 negative cosmological constant $\Lambda=-1$. This theory
 exhibits, among other properties, gravitons; it also
admits black hole solutions \cite{Banados:1992wn}. The
(third order) equations of motion  are
\begin{eqnarray}
G_{\mu\nu}-g_{\mu\nu}+\frac{1}{\mu}C_{\mu\nu}=0,
\label{TMGLambda}
\end{eqnarray}
where $C_{\mu\nu}$ is the Cotton tensor (the analog of the
Weyl tensor in $(3+1)$ dimensions) and the
parameter $\mu$ has the dimension of mass. There has been great recent interest
in what is now known as chiral gravity \cite{Li:2008dq}.

A supersymmetric extension of the equations (\ref{TMGLambda}) have been proposed \cite{Deser}.
The most general supersymmetric solutions \cite{Gibbons:2008vi} read
\begin{eqnarray}
ds^2=\frac{1}{r^2}\left[dr^2+2dtds-f(t)r^{1-\mu}dt^2\right],
\label{solTMG}
\end{eqnarray}
where $\mu^2\not=1$, and $f$ is an arbitrary function of $t$. These solutions
have also been obtained from another point of view, namely by means of a correspondence
between conformal gravity with conformal source and the equations (\ref{TMGLambda}), \cite{AyonBeato:2004fq}.

 When $f$ is constant and  $\mu=3$, we clearly
recover the metric (\ref{BSmetric}).
 For $\mu\neq 3$ the conformal
symmetry is broken (unless the arbitrary function $f$ is chosen as
$f(t)=t^{-3+\mu}$). This particular ``point" $\mu=3$ should exhibit
interesting features worth studying, just like at the chiral point
$\mu=1$.  For $f(t)=a(bt+c)^ {-1}$ and $\mu=2$ we get Dirac's theory
of a variable gravitational constant, conformal with the Kepler
problem \cite{DGH}.

\section{Madelung transcription}

The equations of motion of a perfect fluid in  $(d+1+1)$ dimensions are given by
\begin{eqnarray}
\left\lbrace
\begin{array}{c}
\displaystyle{\partial_t\rho+\bnabla \cdot\left(\rho{\bv} \right)=0}
\\[8pt]
\displaystyle{\rho\left[\partial_t+{\bv} \cdot\bnabla \right]{\bv} =
-\bnabla  p+\rho{\bf f} }
\end{array}
\right.\; ,
\label{eqfluid}
\end{eqnarray}
where $\rho$ is the fluid density, ${\bv} $ the velocity, $p$
the pressure and ${\bf f} $ represents the force density. For a perfect fluid with polytropic equations of motion, i.e.,
$p\propto \rho^{\gamma}$ with polytropic exponent
$\gamma=1+2/(d+1)$,  equations
(\ref{eqfluid}) are Schr\"odinger symmetric \cite{hydro,O'Raifeartaigh:2000mp}.

Decomposing the wave function $\Phi$ of the
modified Schr\"odinger
equation (\ref{modschro}) as
$\Phi=r^{\frac{d+1}{2}}\sqrt{\rho}\,e^{i\Theta}$ yields  the
hydrodynamical system known as the Madelung fluid,
\begin{eqnarray}
&&\partial_t\rho+\frac{1}{m}\bnabla \cdot\left(\rho\bnabla \Theta\right)=0
\\[6pt]
&&\partial_t\Theta+
\frac{1}{2m}\vert\bnabla \Theta\vert^2=-\frac{1}{4m\rho}\left[\frac{1}{2\rho}
\vert\bnabla \rho\vert^2-\Delta\rho\right]-\frac{\alpha}{8mr^2}\nonumber
\label{Madelungfluid}
\end{eqnarray}
where $\alpha=4M^2+(d+3)(d+1)$ and the gradient is w.r.t. $\by=(\bx,r)$. These equations can be interpreted
as the equations of a perfect fluid (\ref{eqfluid}),
whose  motion is irrotational, ${\bv} =\bnabla\Theta$.
It is submitted to an external potential force,
$
{\bf f} =-\bnabla \left(\frac{\alpha}{8mr^2}\right),
$
and has enthalpy
$
\omega=\frac{1}{4m\rho}\left[\frac{1}{2\rho}
\vert\bnabla \rho\vert^2-\Delta\rho\right].
$
For the Madelung system (\ref{Madelungfluid})
the non-relativistic
conformal symmetry is implemented as,
\begin{eqnarray}
&&\Theta(t,\by)\to \Theta(\frac{t}{1-\kappa t}, \frac{\by}{1-\kappa
t})-\kappa\frac{\by^2}{2(1-\kappa t)},
\\[6pt]
&&\rho(t,\by)\to \frac{1}{(1-\kappa t)^{d+1}}\rho(\frac{t}{1-\kappa
t}, \frac{\by}{1-\kappa t}).
\end{eqnarray}

\begin{acknowledgments}
This work has been partially supported by grants 1051084, 1060831 from FONDECYT and by MeceSup Project FSM0605 (Chile). We benefited from correspondence with
Roman Jackiw, Gary Gibbons and Sergej Moroz.
\end{acknowledgments}


\end{document}